\pdfoutput=1
\documentclass[showpacs,pre,twocolumn,superscriptaddress]{revtex4-1}

\usepackage{amsmath,amssymb}
\usepackage{graphicx}
\usepackage{dcolumn}
\usepackage{bm}
\usepackage{color}
\usepackage[normalem]{ulem}

\begin{document}
\title{Mechanically stable nanostructures with desirable characteristic field enhancement factors: a response from scale invariance in electrostatics}

\date{\today}

\author{Thiago A. de Assis}
\address{Instituto de F\'{\i}sica, Universidade Federal da Bahia,
   Campus Universit\'{a}rio da Federa\c c\~ao,
   Rua Bar\~{a}o de Jeremoabo s/n,
40170-115, Salvador, BA, Brazil}
\email{thiagoaa@ufba.br}

\author{Fernando F. Dall'Agnol}
\address{Universidade Federal de Santa Catarina,
 Campus Blumenau,
   Rua Pomerode 710 Salto do Norte, 89065-300,
Blumenau, SC, Brazil}
\email{fernando.dallagnol@ufsc.br}

\begin{abstract}

This work presents an accurate numerical study of the electrostatics of a system formed by individual nanostructures mounted on support substrate tips, which provides a theoretical prototype for applications in field electron emission or for the construction of tips in probe microscopy that requires high resolution. The aim is describe the conditions to produce structures mechanically robust with desirable field enhancement factor (FEF). We modeled a substrate tip with a height $h_1$, radius $r_{1}$ and characteristic FEF $\gamma_1$, and a top nanostructure with a height $h_2$, radius $r_{2}<r_{1}$ and FEF $\gamma_2$, for both hemispheres on post-like structures. The nanostructure mounted on the support substrate tip then has a characteristic FEF, $\gamma_{C}$. Defining the relative difference $\eta_R = (\gamma_{C} - \gamma_{1})/ (\gamma_{3} - \gamma_{1})$, where $\gamma_{3}$ corresponds to the reference FEF for a hemisphere of the post structure with a radius $r_3=r_2$ and height $h_3=h_1 + h_2$, our results show, from a numerical solution of Laplace's equation using a finite element scheme, a scaling $\eta_R = f(u\equiv\lambda\theta^{-1})$, where $\lambda\equiv h_{2}/h_1$ and $\theta = r_1/r_2$. Given a characteristic variable $u_{c}$, for $u \ll u_{c}$, we found a power law $\eta_{R} \sim u^{\kappa}$, with $\kappa \approx 0.55$. For $u \gg u_{c}$, $\eta_{R} \rightarrow 1$, which led to conditions where $\gamma_C \rightarrow \gamma_3$. As a consequence of scale invariance, it is possible to derive a simple expression for $\gamma_C$ and to predict the conditions needed to produce related systems with a desirable FEF that are robust owing to the presence of the substrate tip. Finally, we discuss the validity of Schottky's conjecture (SC) for these systems, showing that, while to obey SC is indicative of scale invariance, the opposite is not necessarily true. This result suggests that a careful analysis must be performed before attributing SC as an origin of giant FEF in experiments.

\end{abstract}

\maketitle

Producing nanostructures that allow one to amplify the applied electric field in their vicinity and which are mechanically stable remains an engineering challenge. This can be observed already a long time ago in the pioneer work by Gomer who discuss a method for growing metal whiskers in a modified field emission tube \cite{Gomer}. In fact, the issue of mechanical stability requires a solution for the degradation and failure of nanostructures that occurs during field electron emission at or near the substrate emitter contact \cite{Bonard} and for the self-mechanical oscillations that occur during field electron emission measurements \cite{Kleshch, Weldon} or from electrostatic interactions \cite{Yasuda}. In particular, a method to study the self-oscillations of a nanostructure mounted on a macroscopic frame requires using a laser beam to excite the sample; subsequently, a second laser beam is then used to register the amplitude of vibrations at a certain point from the object \cite{Eremeyev}.

Applications of these nanostructures mounted on tip devices include carbon nanotubes (CNTs) mounted on a support tip, which can be used as an electron source in a high-resolution electron beam. The latter acquires properties such as a stable emitted current and high brightness \cite{Jonge}. Moreover, due to screening effects \cite{Guihua}, there is a tendency to construct only one structure that protrudes effectively to the substrate tip \cite{Nishijima}. To this end, experiments consisting of mounting multiwalled carbon nanotubes (MWNT) on tungsten probes using, for instance, the dielectrophoresis technique have been performed \cite{Wei}. Alternatively, methods that use chemical vapor deposition (CVD) to grow CNT tips on commercial silicon-cantilever-tips have been developed, leading to MWNTs that are oriented perpendicular to the substrate tip plane \cite{Cheung}. The model we present here can be used to predict the characteristic field enhancement factor (FEF) in these systems.

An open issue related to the experimental and theoretical aspects of the aforementioned studies is how to fabricate nanostructures with a given characteristic FEF that are suitable, for example, for cold field emitter devices that operate at low voltages or for probe microscopes that are mechanically stable, to avoid self-oscillations.
For this purpose, the nanostructure fixation on the tip and its length, which should not be too long, are
part of the requirements for a stable field emission emission, for instance, in probe microscopy experiments conducted with CNT \cite{Shimoi}. However, the disadvantages include a decrease in the characteristic FEF, due to the screening caused by the tip volume, as compared with that of a long nanostructure with a length equal to the combined length of the system (substrate tip + nanostructure) (STN) and a radius equal to that of the thinnest nanostructure. From now on, whenever we mention this specific model, we will indicate it with the initials STN.

In this letter, we address the problem of the effect of the substrate structure (or substrate tip), above which the nanostructure is mounted, on the corresponding characteristic FEF. We show that the scale invariance related to the corresponding electrostatics of this system provides a novel scenario to predict the conditions under which the STN must be developed to produce devices with a desirable FEF and mechanical stability. We also point out a more general scenario for the limits in which Schottky's conjecture (SC) holds \cite{Schottky, Miller, Huang}. As a measure of the sharpness of the system, the actual characteristic FEF (or FEF at the apex of the top protrusion), $\gamma_C$, is defined as

\begin{equation}
\gamma_C = \frac{F_{C}}{F_{M}},
\label{Eq1}
\end{equation}
where $F_{C}$ is the characteristic local barrier field. From the viewpoint of cold field electron emission (CFE) science, $F_{C}$ corresponds to the field defined in the emitter's electrical surface (i.e. ``at the edge" of the surface atoms, and inside the edge of the barrier), which determines the barrier through which the field-emitted electrons tunnel \cite{Forbes1}. In that case, this quantity is typically on the order of a few V/nm for conducting materials and will normally be significantly higher than applied electric field $F_M$.

\begin{figure*}
\includegraphics [width=8.0cm,height=5.5cm] {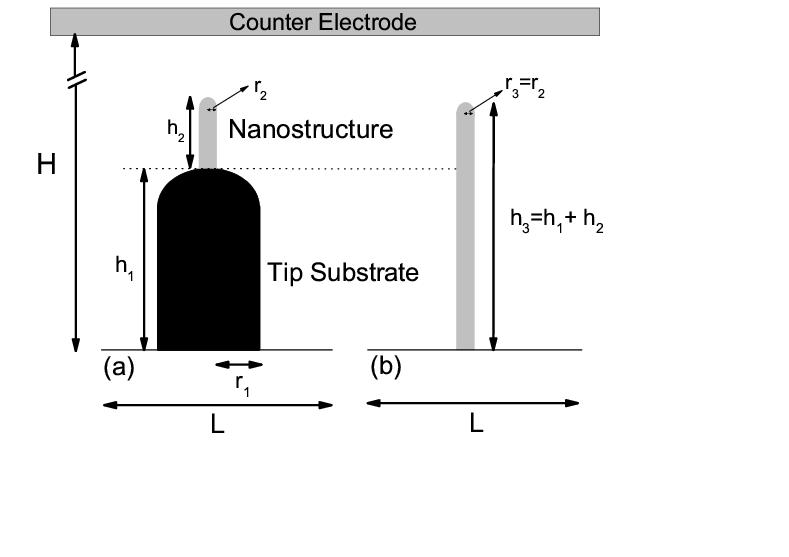}
\caption{(Color online) (a) Two-dimensional representation of the nanostructure mounted on a tip substrate (both hemispheres on post structures) modeled in this work, with an FEF $\gamma_{C}$. The tip substrate has a height $h_1$ and radius $r_1$, and the nanostructure has a height $h_2$ and radius $r_2$; (b) the nanostructure with an FEF $\gamma_3$, height $h_3=h_2 + h_1$ and radius $r_3=r_2$, which is used for comparison. The counter electrode distance and the lateral size of the domain are $H$ and $L$, respectively. In this work, $H = 2\sqrt{2}L$ with $L=5 h_{3}$, and the screening and anode effects on the estimation of $\gamma_C$ are neglected (see text for more details).} \label{Fig1}
\end{figure*}

We modeled a two dimensional axisymmetric system as shown in Fig.\ref{Fig1}(a), formed by two hemispheres placed over cylindrical post structures, which represent the tip substrate and the nanostructure. As isolated and under the applied electric field, the substrate tip has a height $h_1$, radius $r_{1}$ and characteristic FEF $\gamma_{1}$, and the nanostructure used for mounting has a height $h_2$, radius $r_{2} < r_{1}$ and characteristic FEF $\gamma_2$. It is convenient for comparison to define the reference system shown in Fig.\ref{Fig1}(b) with the same radius as the top nanostructure (i.e., $r_3=r_2$), height $h_3 = h_1 + h_2$ and FEF $\gamma_{3}$. Then, the systems shown in Figs. \ref{Fig1}(a) and (b) represent situations of high mechanical stability and of high FEF, respectively. At this point we clarify that desire FEF means that the system STN have $\gamma_C >\gamma_1$ (and $\gamma_C > \gamma_2$), i.e., an additional FEF as compared with a single hemisphere on post system with height $h_1$ ($h_{2}$) and radius $r_1$ ($r_2$), having the possibility of tune the value of $\gamma_C$ as close as possible to $\gamma_3$.

%

\begin{figure*}
\includegraphics [width=12.0cm,height=8.5cm] {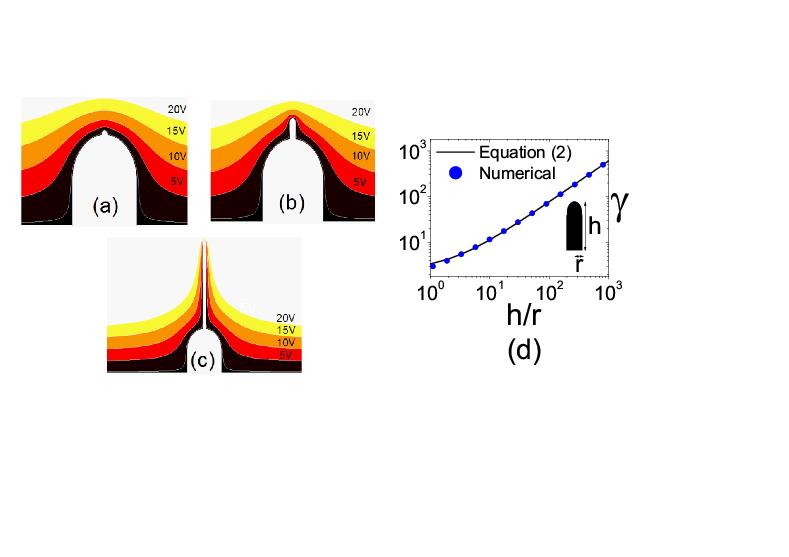}
\includegraphics [width=12.0cm,height=8.5cm] {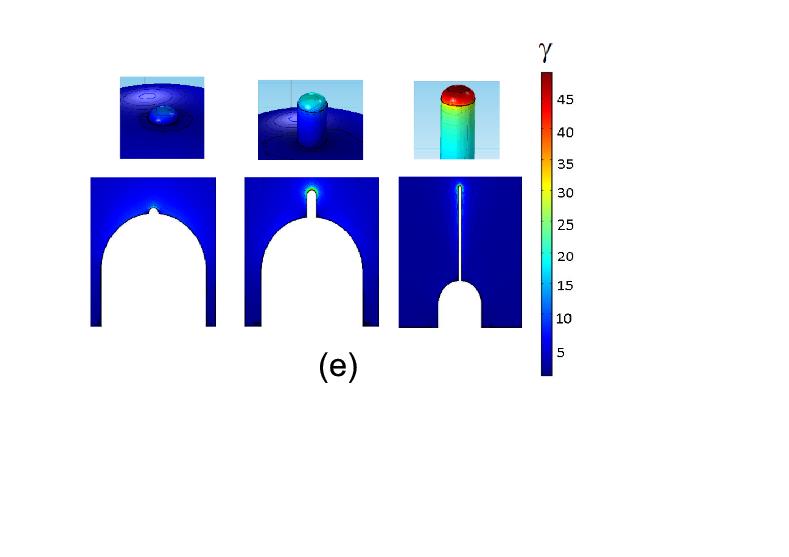}
\caption{(Color online) Equipotential lines calculated numerically for parameter $\lambda \theta^{-1}$ equal to (a) 0.005, (b) 0.025 and (c) 0.3. (d) Comparison between the FEF, as a function of aspect ratio $h/r$, for a hemisphere on a cylindrical post from the numerical solution to Laplace's equation (blue full circles) and from Eq.(\ref{Eq1b}) (black full line). (e) Two-dimensional electric field map for (a) [left], (b) [middle] and (c) [right] configurations (bottom panels) and the corresponding three-dimensional representations (top panels). The FEF color scale is also shown.} \label{Fig2}
\end{figure*}
%

\begin{figure*}
\includegraphics [width=8.0cm,height=5.5cm] {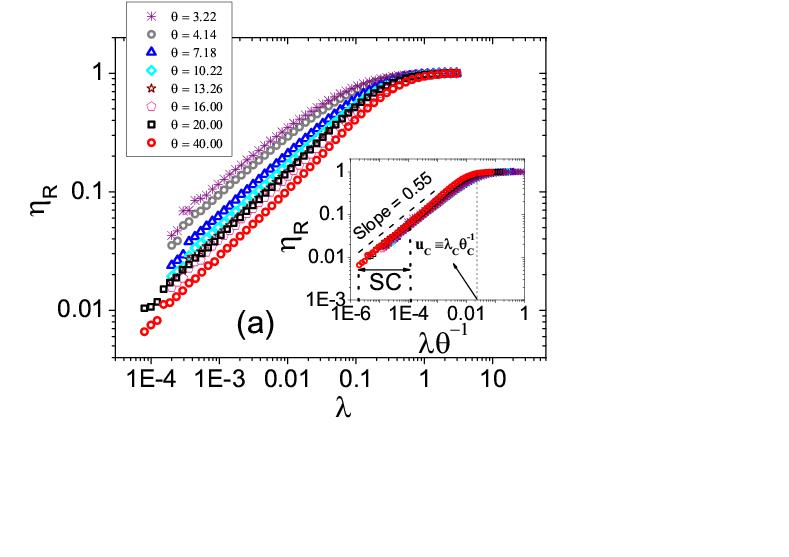}
\includegraphics [width=8.0cm,height=5.5cm] {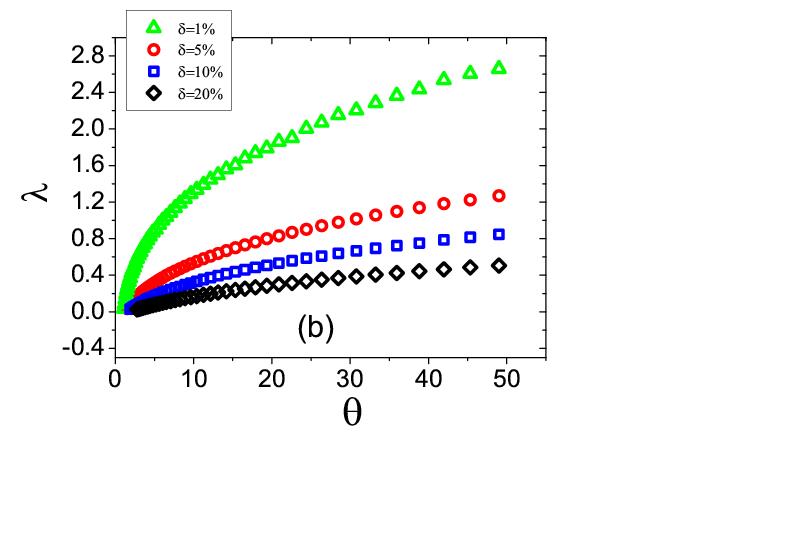}
\caption{(Color online) (a) $\eta_{R}$ [defined in Eq.(\ref{Eq4})] as a function of parameter $\lambda$ [see Eq.(\ref{Eq2})]. In the inset, the results making the scaling transformation $\lambda \rightarrow \lambda\theta^{-1}$, which results in a single curve, are shown. For $u \ll u_{C}$ ($u_{C} \equiv \lambda_{C}\theta_{C}^{-1}$ is the characteristic scaled variable highlighted), $\eta_R \sim u^{\kappa}$. The dashed line has a slope of $0.55$, indicating the approximated value of $\kappa$. The range of $\lambda \theta^{-1}$ that obeys SC is also shown (see text for more details). (b) Behavior of $\lambda$ with respect to $\theta$ for several deviations defined in Eq.(\ref{Eq4b}).} \label{Fig3}
\end{figure*}
%

\begin{figure*}
\includegraphics [width=8.0cm,height=5.5cm] {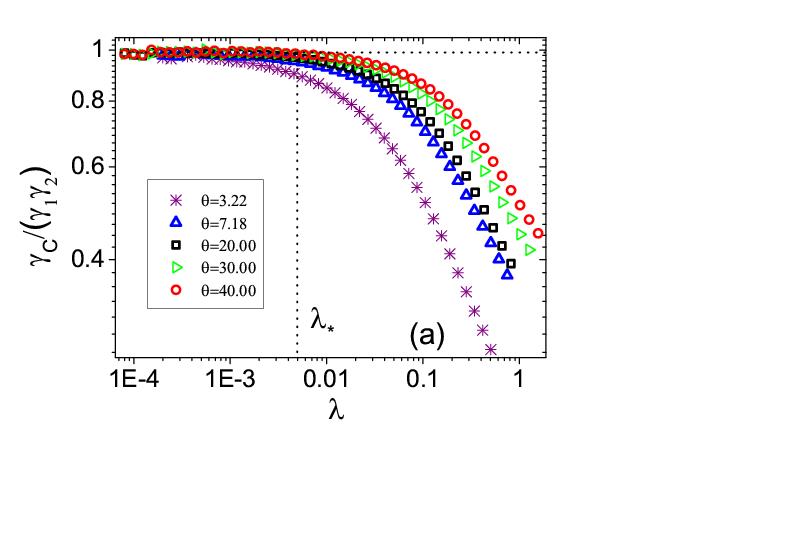}
\includegraphics [width=8.0cm,height=5.5cm] {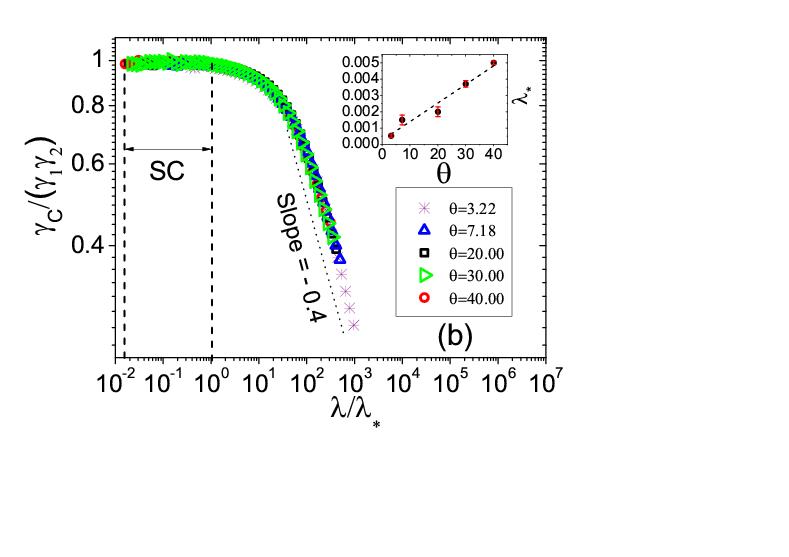}
\caption{(Color online) (a) $\gamma_{C}/\gamma_{1}\gamma_{2}$ as a function of $\lambda$, in the range where $\eta_R \sim u^{0.55}$ (see Fig. \ref{Fig3}) for several values of $\theta$. The representation of $\lambda_{*}$ for $\theta=40$ is also shown (see text for more details). (b) Collapse of the curves shown in (a) under the transformation $\lambda \rightarrow \lambda \lambda_{*}^{-1}$. The dotted line has a slope of $-0.4$. The values of $\lambda_{*}$ as a function of $\theta$ are shown in the inset, with the corresponding error bars. The limits in which SC holds are highlighted.} \label{Fig5}
\end{figure*}
%

\begin{figure*}
\includegraphics [width=8.0cm,height=5.5cm] {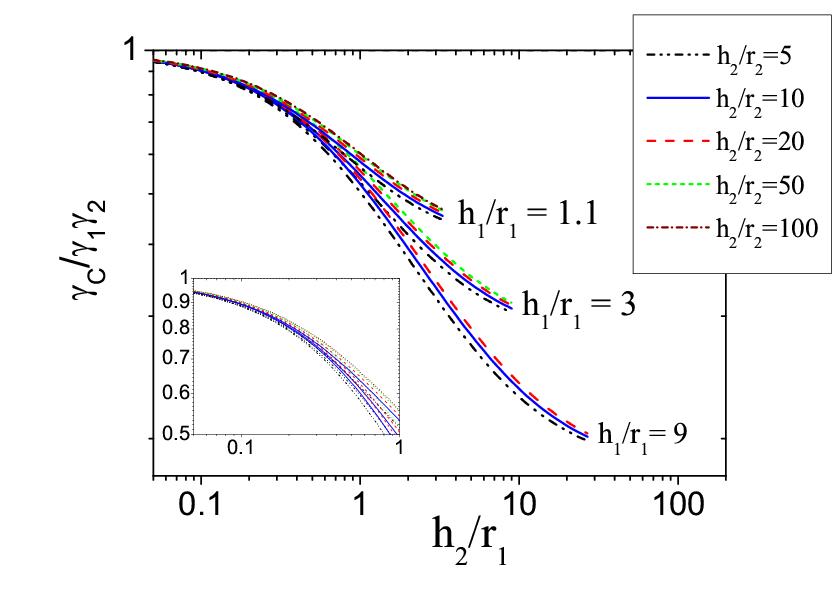}
\caption{(Color online) $\gamma_{C}/\gamma_{1}\gamma_{2}$ as a function of parameter $h_{2}/r_1$, for specific values of $h_{1}/r_{1}$ and $h_{2}/r_{2}$. The inset shows a magnification of main panel for $0 < h_{2}/r_1 \leqslant 1$.} \label{Fig6}
\end{figure*}

In fact, FEF is dependent on the geometrical parameters associated with the device. For a hemisphere on post system, the characteristic FEF for $ 4 \leq h/r \leq 3000$ may be well represented (within $\pm 3\%$) by Edgcombe's formula \cite{Forbes1}

\begin{equation}
\gamma_{n} \approx 1.2 \left(2.15 + \frac{h_n}{r_n}\right)^{0.9},
\label{Eq1b}
\end{equation}
where $n=1,2,3$. The solution to this problem is far from trivial, and no exact analytical solution is known. Therefore, the electrostatics of a nanostructure on a substrate tip, despite having potential applications, are still considered an open theoretical issue. For a diode system such as that used in this work, the main parameters that affect the FEF are expected to be the gap between the device and the counter electrode, $H$ \cite{Forbes1,Smith,Zheng} and the following dimensionless parameters

\begin{equation}
\lambda \equiv \frac{h_2}{h_1},
\label{Eq2}
\end{equation}
and
\begin{equation}
\theta \equiv \frac{r_1}{r_2} > 1.
\label{Eq3}
\end{equation}
To check the validity of our procedure, we numerically solved Laplace's equation in a two-dimensional axisymmetric domain within the range $ 4 \leq h/r \leq 1000$, and we assumed the tip substrate and the top nanostructure as conductors. The electric potential distribution on the integration domain was calculated using the finite elements method (using the COMSOL® v4.3b software package), thus allowing the calculation of the electric field distribution over the device. In this work, we use a domain with the lateral size $L=5 h_{3}$ such that the screening from neighbors is considered negligible (the nanostructures can be considered isolated) \cite{FAD}. The electric potential $\Phi^{A}\neq 0$ of the counter electrode at the top boundary guarantees an electric field intensity that is equal to $F_M$ at the boundary. Moreover, the emitter surface and the bottom boundary of the cell domain are assumed to be grounded ($\Phi^{S}$=0). We have also considered $H/L=2\sqrt{2}$, which satisfies the condition for setting the counter electrode distance without affecting the calculated FEFs. In Figs.\ref{Fig2} (a), (b) and (c), we show the equipotential lines for systems similar to that shown in Fig.\ref{Fig1}(a) considering $\lambda \theta^{-1}$ = 0.005, 0.025 and 0.3, respectively. Whereas for $\lambda \theta^{-1}$=0.005, the form of the equipotential lines is predominantly due to the tip substrate geometry (with a small influence from the nanostructure on the tip for $\Phi$ = 5 V), for $\lambda \theta^{-1}$=0.3, the effect of the tip substrate on the form of the equipotential lines is negligible at the top of the nanostructure, and $\gamma_C$ is expected to be close to $\gamma_3$. This limiting case represents the situation in which the tip substrate does not approximately affect the value of the characteristic FEF. To demonstrate the accuracy of our calculations, we present in Fig.\ref{Fig2} (d) a comparison between the results from Eq.(\ref{Eq1b}) and those from our numerical calculations for a problem of a single hemisphere on a cylindrical post. Excellent agreement is achieved for experimental plausible range, i.e. for $100 \leq h/r \leq 1000$, a deviation less than 2\% has been found. We also found for $h/r=1$, $\gamma_{n} \approx 2.98$, characterizing a deviation of approximately $0.7$\% from theoretical predictions ($\gamma=3$). Moreover, the simple approximation $\gamma_{n} \approx 0.7 h_{n}/r_{n}$ gives results about 10\% too high over the same range. In contrast, Ref.\cite{Roveri} has obtained a small shifted curve compared with our results; thus, the claim that their simulation results have converged into other analytical expressions [different from Eq.(\ref{Eq1b})] is not clear. In Fig.\ref{Fig2}(e), we show the corresponding two-dimensional electric field map for cases (a), (b) and (c) (bottom panels) and the corresponding three-dimensional representation (top panels). The FEF in the color scale is also shown for $1 \leq \gamma \leq 50$. To quantitatively evaluate the effect of substrate tip on $\gamma_C$, we define the relative difference as

\begin{equation}
\eta_R \equiv \left(\frac{\gamma_C - \gamma_1}{\gamma_3 - \gamma_1}\right),
\label{Eq4}
\end{equation}
such that $0 \leq \eta_R \leq 1$, with lower and upper limit values obtained for $\gamma_C \rightarrow \gamma_1$ and for $\gamma_C \rightarrow \gamma_3$, respectively. In Fig. \ref{Fig3}(a), we show the results of $\eta_R$ as a function of parameter $\lambda$ for different values of parameter $\theta$. We observe a clear power law behavior, which indicates that a slow increase in $\eta_R$ occurs as $\lambda$ increases in this range. This result shows that $\gamma_C$ slowly approaches $\gamma_3$ as $h_2$ increases, for a fixed $h_1$, in this regime. Moreover, for a fixed $\lambda \ll 1$, we observed a tendency for $\eta_R$ to decrease as $\theta$ increases, showing that large radius $r_1$ may produce a screening effect, decreasing the value of $\gamma_C$. Interestingly, we can see in the inset of Fig.\ref{Fig3}(a) that all curves shown in the principal panel collapse into a single curve under the transformation $\lambda \rightarrow \lambda\theta^{-1}$. This result is a signature of scale invariance in these systems. If we define the variable $u\equiv\lambda\theta^{-1}$, our results show that $\eta_R = f(u)$; thus, for $u \ll u_{c}$, where $u_{c}$ is a characteristic scaled variable, $\eta_R \sim u^{\kappa}$ with $\kappa \approx 0.55$. In contrast, for $u \gg u_{c}$, $\eta_R \approx 1$, which means that, in this regime, the substrate tip does not influence the $\gamma_C$ values of the STN system, which are expected to be close to $\gamma_3$. Therefore, we propose the follow ansatz:

\[ \eta_R = f(u) \sim \left\{ \begin{array}{ll}
         u^{\kappa} & \mbox{if $u \ll u_{c}$};\\
         1 & \mbox{if $u \gg u_{c}$}.\end{array} \right. \]

Using Eqs.(\ref{Eq1b}), (\ref{Eq4}), and the ansatz above, we propose the following equation for $\gamma_C$ for our STN:

\begin{equation}
\gamma_C \approx \eta_R \left(\gamma_3 - \gamma_1\right) + \gamma_{1}.
\label{Eq4a}
\end{equation}
Equation (\ref{Eq4a}) and the ansatz for $\eta_R$ allow one to predict the geometrical parameters of a nanostructure mounted on tip substrate to produce a robust STN system with FEF $\gamma_C = (1-\delta)\gamma_3$, where $\delta$ is the relative deviation. The relative deviation can be derived as a function of $\eta_R$ as follows:

\begin{equation}
\delta = (1-\eta_R)\left(1 - \frac{\gamma_1}{\gamma_3}\right).
\label{Eq4b}
\end{equation}
Equation (\ref{Eq4b}) has a clearer interpretation as the fraction between the desired $\gamma_3$ with actual $\gamma_C$. Therefore, $\lambda$ and $\theta$ can be related to provide $\delta$ with a given tolerance. Fig.\ref{Fig3}(b) shows, for a given value of $\delta$, how $\lambda$ must behave with respect to $\theta$. A conclusion is that for $\delta = 10\%$ (which is considered a reasonable limit for experimental precision), the effect of $\theta$ on $\lambda$ is small, thereby suggesting that a significant range of $\theta$ may be used to produce approximately the same value of $\gamma_C$, deviating only $10\%$ from corresponding $\gamma_3$.

Next, we address the validity of the Schottky Conjecture (SC) \cite{Schottky} in this system, i.e., as the FEF of the microprotrusion on top of a macroprotrusion is dominated by the product of the individual protrusions' FEFs (in our case $\gamma_C = \gamma_1 \gamma_2$). This problem has been addressed analytically by Miller \textit{et al.} using conformal mapping for several rectilinear geometries in two dimensional systems, in contrast with our STN system, which considers rotational symmetric hemispheres on posts. They found that a significant deviation from Schottky's product rule occurs almost exclusively when the half-width of the macroprotrusion is less than the height of the microprotrusion \cite{Miller}. In a very recent work, Jensen \textit{et. al} \cite{Jensen} provided a mathematical proof for protrusions of conical/ellipsoidal shapes using a point charge model. Interestingly, they found that Schottky's conjecture remains valid even for similar dimensions of protrusion and base structure. On the experimental side, Huang \textit{et al.} \cite{Huang} demonstrated strong field emission from carbon nanotubes grown on carbon cloth. Their results were justified claiming that the FEFs of emitters with a multistage result from the product of the individual FEFs of the individual stages.

In Fig. \ref{Fig5}(a), we show the behavior of $\gamma_{C}/(\gamma_{1}\gamma_{2})$ as a function of $\lambda$ for the region where $\eta_R$ scales as $(\lambda \theta^{-1})^{0.55}$ [see Fig. \ref{Fig3}(a)]. The results show that the validity of $\gamma_{C} \approx \gamma_{1}\gamma_{2}$ increases with $\theta$, which makes it possible to define a characteristic variable $\lambda_{*}$. We estimate $\lambda_*$ where the condition $\gamma_{C}/ (\gamma_{1}\gamma_{2}) \approx 0.9$ is satisfied, and we evaluate how $\lambda_*$ scales with $\theta$. The results shown in Fig. \ref{Fig5}(b) show a simple linear relationship of $\lambda_* \approx 10^{-4} \theta$ (at least for $3 \lesssim \theta \lesssim 40$ - see inset). A different dependence between $\lambda_*$ and $\theta$ is not discarded for high values of $\theta$. Moreover, under transformation $\lambda \rightarrow \lambda \lambda_{*}^{-1}$, all curves shown in Fig. \ref{Fig5}(a) exhibit an excellent collapse, which suggests that

\begin{equation}
 \gamma_{C} = g\left( \lambda \theta^{-1}\right) \gamma_{1}\gamma_{2},
 \label{Eq6}
\end{equation}
where $g\left( \lambda \theta^{-1}\right)$ is a correction, in the form of a scaling function, that generalizes the SC. Equation (\ref{Eq6}) and numerical results from Fig.\ref{Fig5}(b) allow one to predict that the range in which SC works is restricted to $10^{4}\lambda\theta^{-1}\approx\lambda \lambda_{*}^{-1} \lesssim 1$. This restriction implies $g\left( u \right) = 1$ for $u \lesssim 10^{-4}$ [as indicated in the inset of Fig.\ref{Fig3}(a)]. A clear regime in which $g\left( u \right) \sim u^{-\nu}$, with $\nu \approx 0.4$, is also identified. Therefore, $g(u)$ assumes the following form:

\[ g(u) \sim \left\{ \begin{array}{lll}
         1 & \mbox{if $\lambda \ll \lambda_{*}$};\\
         u^{-\nu} & \mbox{if $\lambda_{*} \ll \lambda \ll \lambda_{c}$};\\
         \gamma_1^{-1} & \mbox{if $\lambda \gg \lambda_{c}$}.\end{array} \right. \].

Interestingly, SC does not encompass all intervals in which the scaling $\eta_R \sim u^{0.55}$ works but is rather valid only in a narrow range. More specifically, it is valid in situations in which the main interest is to produce structures that are mechanically stable but do not have promising FEFs. In these situations, careful analysis is required to attribute SC as an explanation for the origin of the giant FEF reported in the experiments. Lack of emission orthodoxy may sometimes be an alternative explanation \cite{Forbes2013}. We stress that as $\eta_R \approx 1$ (i.e., $\lambda \gg \lambda_{c}$), $\gamma_{C}/(\gamma_{1}\gamma_{2})$ is expected converge to $\gamma_{1}^{-1}$ (data not shown).

Finally, with the aim to compare our results with those of Refs. \cite{Miller} and \cite{Jensen}, we show in Fig. \ref{Fig6} the behavior of $\gamma_{C}/(\gamma_{1}\gamma_{2})$ as a function of parameter $h_{2}/r_1$, for specific values of $h_{1}/r_{1}$ and $h_{2}/r_{2}$. The results suggest that SC works well in our system if $h_{2} \ll 0.1 r_1$. As compared with the results from geometries used in Ref.\cite{Miller}, it's possible to observe that in the range $ 0< h_2/r_1 \leq 1$, $\gamma_{C}/(\gamma_{1}\gamma_{2})$ exhibit a more pronounced decay as $h_2/r_1$ increases (see inset), suggesting a more restricted interval where SC works. Moreover, considering our simulation space, no evidence of SC has been found as similar dimensions of protrusion and base structure are considered. As already pointed, this feature has been found analytically in Ref.\cite{Jensen}, but for conical field emitters using point charge model.

In summary, we have proposed a novel scaling ansatz for electrostatic systems that allows prediction of the geometrical conditions of the tip substrate, with the main purpose of producing devices that are mechanically stable and have a desirable FEF. For systems formed by nanostrucutures mounted on a substrate tip (both hemispheres on post-like structures), the results show that the influence of the substrate on the FEF of a mechanically stable system may be explained by scale invariance arguments. We found that $\eta_R$ is a function of variable $u\equiv\lambda \theta^{-1}$, the latter of which is related to the geometry of the tip and top nanostructure. Finally, our results show that SC is respected for a restricted interval of the scaling variable $u$, thus indicating that while to obey SC is indicative of scale invariance, the opposite is not necessarily true. This result provides important clues for the design of nanostructures on tip emitters for electronic applications that are beyond the limits of SC.

\section*{Acknowledgements}

The authors acknowledge the financial support of the Brazilian agency CNPq.

\vspace{1.0cm}

%

\end{document}